\documentclass{moriond}

\def\be{\begin{equation}}
\def\ee{\end{equation}}
\def\bea{\begin{eqnarray}}
\def\eea{\end{eqnarray}}

\newcommand\POWHEG{{\tt POWHEG}}
\newcommand\MCatNLO{{\tt MC@NLO}}
\newcommand\MiNLO{{\tt MiNLO}}
\def\sss{\scriptscriptstyle}
\newcommand\as{\alpha_{\sss\rm S}}
\newcommand\Phirad{\Phi_{\mathrm{rad}}}
\def\nn{\nonumber}
\usepackage{amssymb}
\usepackage{amsmath}
\newcommand{\noun}[1]{\textsc{#1}}
\newcommand{\HNNLO}{\noun{Hnnlo}}
\newcommand{\HQT}{\noun{HqT}}
\newcommand{\JETVHETO}{\noun{JetVHeto}}
\newcommand{\Pythia}{\noun{Pythia~6}}

\newcommand{\Photo}{}

\begin{document}
\vspace*{4cm}
\title{HIGGS PRODUCTION AT NNLOPS}

\author{E.~RE}

\address{Rudolf Peierls Centre for Theoretical Physics, Department
  of Physics, University of Oxford,\\
  Oxford, OX1 3NP, United Kingdom}

\maketitle\abstracts{We describe the method used to build a simulation
  of Higgs boson production accurate at next-to-next-to-leading order
  and matched to a parton shower. The adopted procedure makes use of a
  combination of the \POWHEG{} and \MiNLO{} methods. We also use
  results from \HNNLO{} as final input to reach the claimed
  accuracy. Results for typical observables are shown.}

%% We describe the method used to build a simulation of Higgs boson
%% production accurate at next-to-next-to-leading order and matched to
%% a parton shower and show results for typical observables. The
%% adopted procedure makes use of a combination of the POWHEG and
%% MiNLO methods. We use results from Hnnlo as final input to reach
%% the claimed accuracy.

\section{Introduction}

During the last decade a major research effort in the Monte Carlo
community has been devoted to the development of NLOPS tools,
\emph{i.e.} tools that allow a matching of next-to-leading order (NLO)
computations with parton showers (PS), thereby bringing NLO accuracy
into standard Monte Carlo event
generators~\cite{Buckley:2011ms}. Among many proposals, there are
currently two well-established NLOPS approaches, namely
\POWHEG{}~\cite{Nason:2004rx,Frixione:2007vw} and
\MCatNLO{}~\cite{Frixione:2002ik}, which have now become the methods
of choice used by experimental collaborations in many analyses being
carried out at the LHC.

Despite in general NLO accuracy is enough for the majority of
processes studied at the LHC, it is known that for some of them the
inclusion of next-to-next-to-leading order (NNLO) effects is
necessary: this is the case when the experimental accuracy demands
$\mathcal{O} (1\%) $ precision in theoretical predictions, or when
NNLO effects are large. Paramount examples of these 2 situations are
Drell-Yan and (gluon-fusion-initiated) Higgs production,
respectively. In these cases, it is clearly desirable to include NNLO
corrections into Monte Carlo programs, if one wants to have a
simulation tool which is flexible and accurate enough at the same
time.

A NNLOPS simulation was achieved recently for Higgs
production~\cite{Hamilton:2013fea}. In this document the theoretical
ingredients (namely the \POWHEG{} and \MiNLO{} approaches) underlying
this result are quickly summarised, the method used to combine them is
outlined, and selected results are shown.

\section{Higgs production at NNLOPS}
\subsection{\POWHEG{}}
The \POWHEG{} method is a prescription to match NLO calculations with
parton shower generators avoiding double counting of real emissions
and virtual corrections.  In the \POWHEG{} formalism, the generation
of the hardest emission is performed first, according to the
distribution given by
\begin{equation}
\label{eq:master}
d\sigma=\bar{B}\left(\Phi_{B}\right)\,d\Phi_{B}\,\left[\Delta_{R}\left(p_{T}^{\min}\right)+
\frac{R\left(\Phi_{R}\right)}{B\left(\Phi_{B}\right)}\,\Delta_{R}\left(k_{T}\left(\Phi_{R}\right)\right)\,d\Phi_{\mathrm{rad}}\right]\,,
\end{equation}
where $B\left(\Phi_{B}\right)$ is the leading order term,
\begin{equation}
\label{eq:bbar}
\bar{B}\left(\Phi_{B}\right)=B\left(\Phi_{B}\right)+
\left[V\left(\Phi_{B}\right)+\int d\Phi_{\mathrm{rad}}\, R\left(\Phi_{R}\right)\right]
\end{equation}
is the NLO differential cross section integrated on the radiation
variables while keeping the Born kinematics fixed
($V\left(\Phi_{B}\right)$ and $R\left(\Phi_{R}\right)$ stand
respectively for the virtual and the real corrections), and
%\begin{equation}
$\Delta_{R}\left(p_{T}\right)=\exp\left[-\int d\Phi_{\mathrm{rad}}\,\frac{R\left(\Phi_{R}\right)}{B\left(\Phi_{B}\right)}\,\theta\left(k_{T}\left(\Phi_{R}\right)-p_{T}\right)\right]\,$
%\end{equation}
is the \POWHEG{} Sudakov form factor. With $k_T\left(\Phi_{R}\right)$
we denote the transverse momentum of the emitted particle off a
Born-like kinematics $\Phi_B$.  It can be shown that by showering the
partonic events generated according to eq.~(\ref{eq:master}), one
obtains NLOPS-accurate results, \emph{i.e.} Sudakov suppression close
to the soft-collinear regions, LO accuracy in the regions where the
\POWHEG{} emission is widely-separated from the other coloured
particles in $\Phi_B$, and, crucially, NLO accuracy for inclusive
observables.

From the NLOPS-matching point of view, the more challenging processes
currently described with this approach are $2\to 3$ and $2\to 4$
processes, with at most 2 light jets at
LO~\cite{Campbell:2012am,Re:2012zi,Jager:2012xk,Jager:2013mu,Jager:2013iza}. One
should notice that when one or more jets are present at LO (as in the
$H+1$ jet case), the $\bar{B}$ function needs to be regulated from the
divergences arising when jets in the LO kinematics become
unresolved~\cite{Alioli:2010qp}: as a consequence, a \POWHEG{}
simulation of $H+1$ jet cannot be used to describe Higgs production
observables that are fully inclusive over QCD radiation.

\subsection{\MiNLO{}}
The \MiNLO{} procedure~\cite{Hamilton:2012np} was originally
introduced as a prescription to a-priori choose the renormalisation
($\mu_R$) and factorisation ($\mu_F$) scales in multileg NLO
computations: since these computations can probe kinematical regimes
involving several different scales, the choice of $\mu_R$ and $\mu_F$
is indeed ambiguous, and the \MiNLO{} method addresses this issue by
consistently including CKKW-like
corrections~\cite{Catani:2001cc,Lonnblad:2001iq} into a standard NLO
computation.  By clustering with a $k_T$-measure the momenta of each
phase-space point sampled, one can define the ``most-probable''
branching history that would have produced such a kinematics:
similarly to what is done in parton showers, the argument of each
power of $\as$ is then found from the transverse momentum of the
splitting occurring at each nodal point of the skeleton built from
clustering, and a prescription for $\mu_F$ is given as well. The
result is also corrected by means of Sudakov form factors (called
\MiNLO{}-Sudakov FF's in the following) associated to internal lines,
accounting for the large logarithms that arise when the clustered
event contains well separated scales.

Because of the presence of \MiNLO{}-Sudakov FF's associated to the
Born-like kinematics, the integration over the full phase space
$\Phi_B$ can be performed without generation cuts, yielding finite
results also when jets in the LO kinematics become unresolved. As a
consequence, the \MiNLO{} procedure can be used within the \POWHEG{}
formalism to regulate the $\bar{B}$ function for processes involving
jets at
LO~\cite{Hamilton:2012np,Hamilton:2012rf,Campbell:2013vha,Luisoni:2013cuh},
without using external cuts or variants thereof.  In particular, in
the $H+1$ jet case, the master formula for generating the hardest
emission contains the following \MiNLO{}-improved $\bar{B}$ function
\begin{eqnarray}
  \label{eq:bbarH1jMiNLO}
  \bar{B}_{\,\tt HJ-MiNLO}&=&\as^2(M_H) \as(q_T) \Delta^2_g(q_T,M_H)\\
  &\times&\Big{[}
  B ( 1-2\Delta^{(1)}_g(q_T,M_H) ) + 
  \as V(\bar{\mu}_R) +\as \int d\Phirad R
  \Big{]}\nn\,,
\end{eqnarray}
to be contrasted with the normal \POWHEG{} $\bar{B}$ function, that
would read for this process
\begin{equation}
  \label{eq:bbarH1j}
  \bar{B}_{\,\tt HJ}=\as^3(\mu_R) \Big{[} B + \as V(\mu_R) + \as \int d\Phirad R \Big{]}\,.
\end{equation}
In eq.~(\ref{eq:bbarH1jMiNLO}) $q_T$ and $M_H$ are the Higgs
transverse momentum and virtuality, $\bar{\mu}_R$ is set to $(M_H^2
q_T)^{1/3}$ in accordance with the \MiNLO{} prescription and
$\Delta_g(q_T,Q)=\exp\Big{\{}-\int_{q^2_T}^{Q^2}\frac{dq^2}{q^2}\frac{\as(q^2)}{2\pi}
\Big{[} A_{g}\log\frac{Q^2}{q^2} + B_{g} \Big{]}\Big{\}}$ is the
\MiNLO{}-Sudakov FF associated to the jet present at LO. The term in
brackets multiplying $B$ contains $\Delta_{g}^{(1)}(q_T,Q)$, that is
the $\mathcal{O} (\as)$ expansion of $\Delta_g$.

In ref.~\cite{Hamilton:2012rf} it was found that not only
eq.~(\ref{eq:bbarH1jMiNLO}) allows to integrate over the full phase
space associated with the ``LO'' jet, but also that, with relatively
minor improvements, the result so-obtained is formally NLO accurate
also for fully-inclusive observables.

\subsection{NNLOPS}
The $H+1$ jet \POWHEG{} implementation enhanced with the improved
\MiNLO{} procedure previously outlined can be used to reach NNLOPS
accuracy. In fact, since such a simulation gives a NLO-accurate
prediction of the Higgs rapidity ($y$), then the function $W(y)$,
defined as
\begin{equation}
  \label{eq:wy}
  W(y)=\frac{(d\sigma/dy)_{\rm NNLO}}
  {(d\sigma/dy)_{\tt HJ-MiNLO}}\,,
  % =\frac{c_2\as^2 + c_3\as^3 + c_4\as^4}{c_2\as^2 + c_3\as^3 +
  %   d_4\as^4} \simeq 1 + \frac{c_4 - d_4}{c_2}{\as^2} +
  % \mathcal{O}(\as^3)
\end{equation}
can be used to reweight each {\tt HJ-MiNLO}-generated event, thereby
obtaining a NNLOPS simulation of inclusive Higgs production.  By
NNLOPS we mean a fully-exclusive Monte Carlo simulation of
Higgs-production which is NNLO accurate for fully-inclusive
observables, as well as LO (NLO) accurate for $H+2 (1)$ jet
observables~\cite{Hamilton:2012rf,Hamilton:2013fea}. Since we are
reweighting the events with $W(y)$, the Higgs rapidity is NNLO
accurate by construction, whereas the NLO accuracy of the 1-jet
region, inherited from the underlying {\tt HJ-MiNLO} simulation, is
not spoilt, because the first non-controlled terms in the whole
simulation are $\mathcal{O}(\as^5)$: this follows from the fact that
$W(y)=1+ \mathcal{O}(\as^2)$, as can be seen expanding numerator and
denominator in eq.~(\ref{eq:wy}).

In ref.~\cite{Hamilton:2013fea} the following generalisation of
eq.~(\ref{eq:wy}) was used:
\begin{equation}
  \label{eq:wypaper}
  W(y,p_T)=h(p_T)\frac{\int d\sigma_{\rm NNLO}\delta(y-y(\Phi)) - \int d\sigma_{\tt HJ-MiNLO}^B \delta(y-y( \Phi)) }{\int d\sigma_{\tt HJ-MiNLO}^A \delta(y-y(\Phi)) } + (1-h(p_T))\,,
\end{equation}
where we have split the {\tt HJ-MiNLO} differential cross section
among $d\sigma^A = d\sigma\ h(p_T)$ and
$d\sigma^B=d\sigma\ (1-h(p_T))$, with $h(p_T)=\frac{(\beta
  m_H)^2}{(\beta m_H)^2+p_T^2}$. The profiling function $h$ controls
where the NLO-to-NNLO correction is spread: as
$2^{\mbox{\scriptsize{nd}}}$ argument of $W$ the transverse momentum
of the leading jet was used, and we have chosen $\beta=1/2$, which
implies that the NNLO correcting factor $W$ is effectively applied in
the region $p_T\lesssim m_H/2$.

\subsection{Results}
In our simulation, the central value for $d\sigma_{\rm NNLO}$ was
obtained with \HNNLO{}~\cite{Catani:2007vq,Grazzini:2008tf}, setting
$\mu_R=\mu_F=m_H/2$. We refer to ref.~\cite{Hamilton:2013fea} for
details on how scales were varied to obtain uncertainty bands.

\begin{figure}[!htb]
  \begin{minipage}{0.5\linewidth}
    \centerline{\includegraphics[scale=0.6]{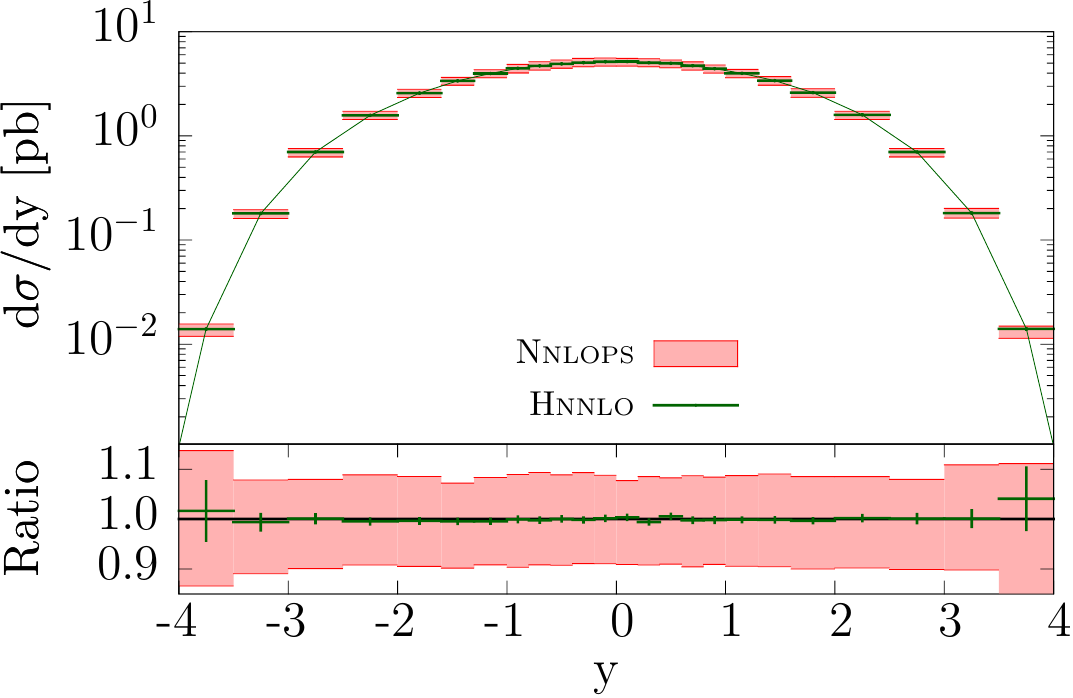}}
  \end{minipage}
  \begin{minipage}{0.5\linewidth}
    \centerline{\includegraphics[scale=0.6]{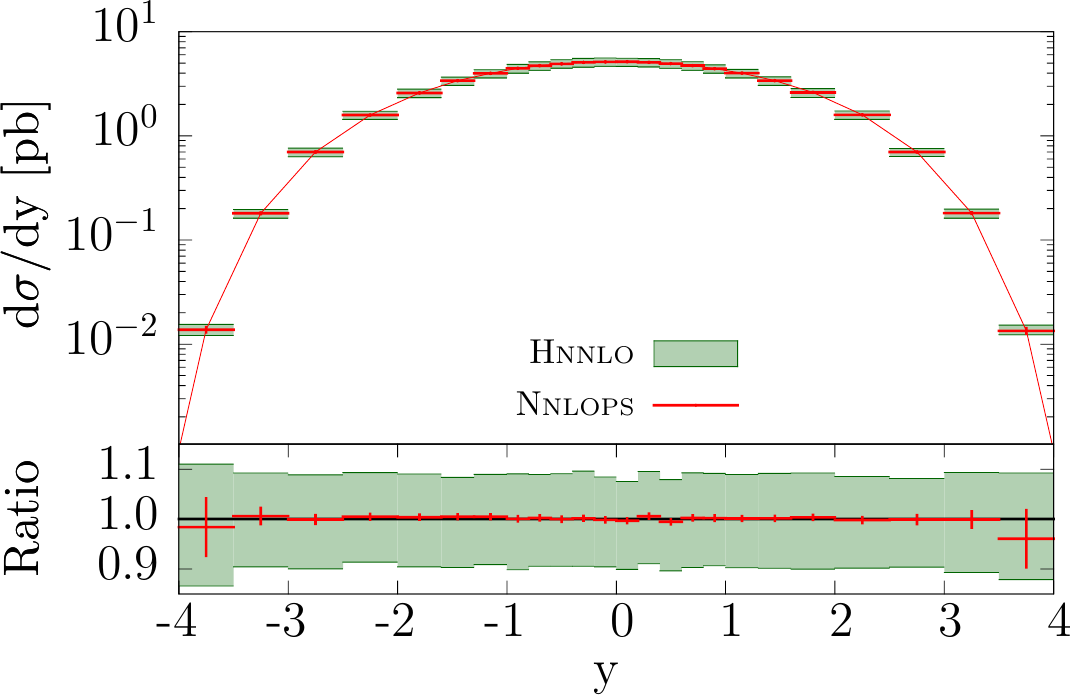}}
  \end{minipage}
  \caption{Comparison of the NNLOPS (red) and \HNNLO{} (green) results
    for the Higgs fully inclusive rapidity distribution. On the left
    (right) plot only the NNLOPS (\HNNLO{}) uncertainty is displayed.
    The lower left (right) panel shows the ratio with respect to the
    NNLOPS (\HNNLO{}) prediction obtained with its central scale
    choice.}
  \label{fig:y}
\end{figure}

A comparison between our NNLOPS simulation and \HNNLO{} is shown in
fig.~\ref{fig:y} (partonic events were showered with
\Pythia{}~\cite{Sjostrand:2006za}): as expected, the NNLOPS simulation
reproduces extremely well the NNLO results for the Higgs rapidity both
in the central value and in the uncertainty band obtained by scale
variation.

%% \begin{figure}[!htb]
%%   \begin{center}
%%     \includegraphics[scale=0.6]{plots/NNLOPS/fig_1/fig1L.pdf}~
%%     \includegraphics[scale=0.6]{plots/NNLOPS/fig_1/fig1R.pdf}
%%     \caption{Comparison of the NNLOPS (red) and \HNNLO{} (green) results
%%       for the Higgs fully inclusive rapidity distribution. On the left
%%       (right) plot only the NNLOPS (\HNNLO{}) uncertainty is displayed.
%%       The lower left (right) panel shows the ratio with respect to the
%%       NNLOPS (\HNNLO{}) prediction obtained with its central scale
%%       choice.}
%%     \label{fig:y}
%%   \end{center}
%% \end{figure}

\begin{figure}[!htb]
  \begin{minipage}{0.5\linewidth}
    \centerline{\includegraphics[scale=0.6]{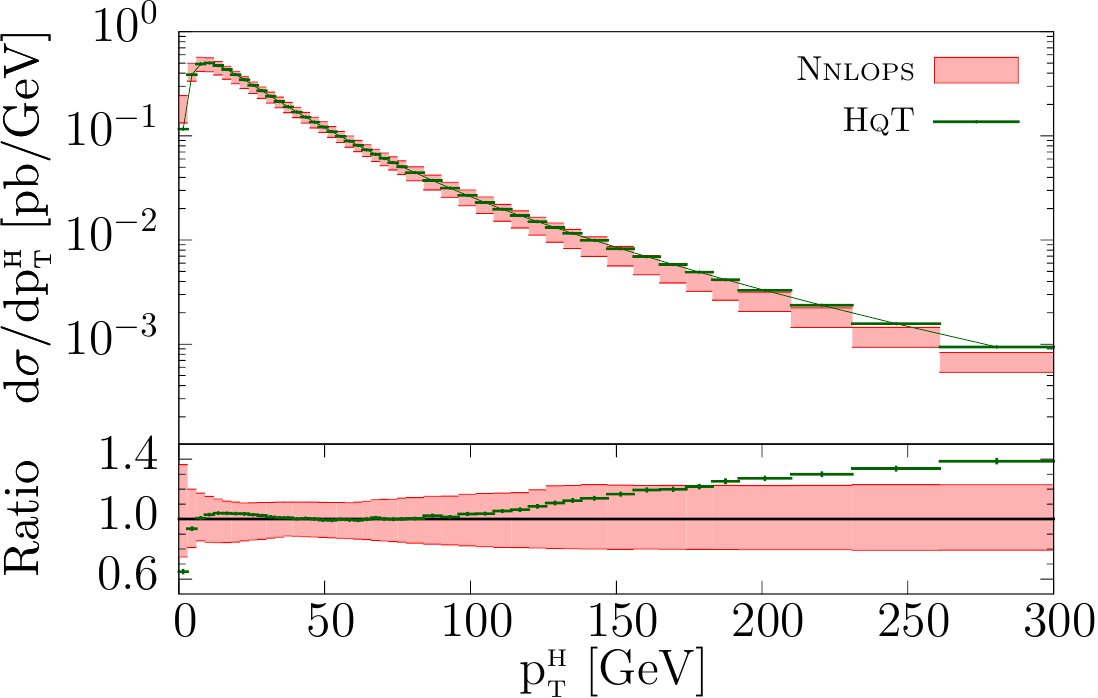}}
  \end{minipage}
  \begin{minipage}{0.5\linewidth}
    \centerline{\includegraphics[scale=0.6]{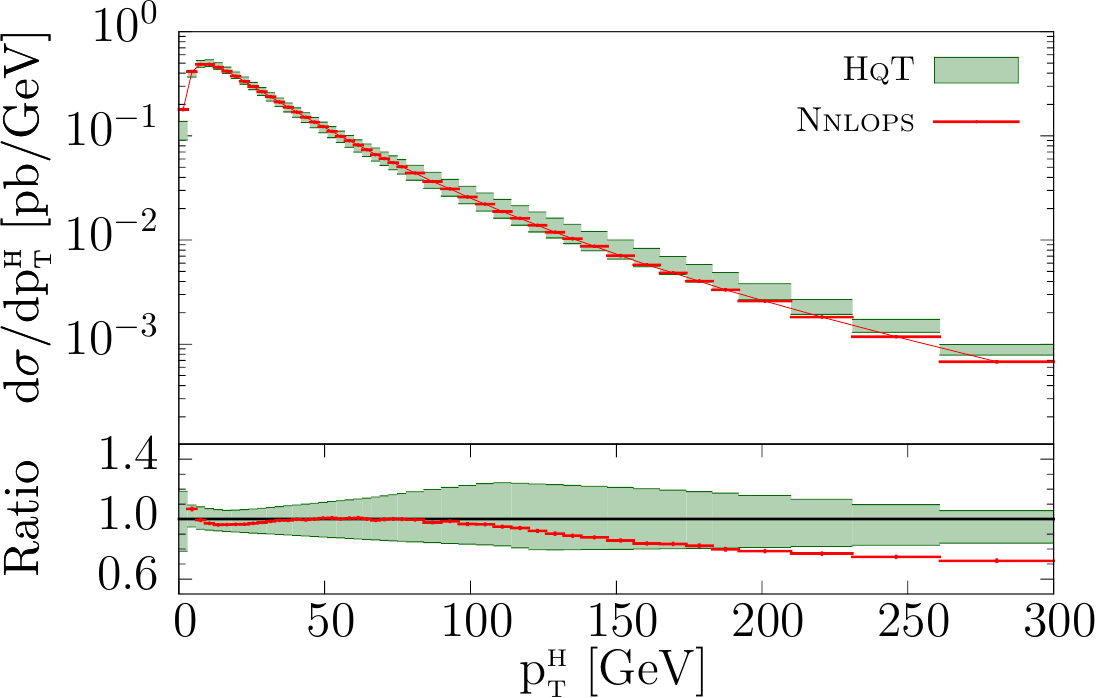}}
  \end{minipage}
    \caption{Comparison of the NNLOPS (red) with the NNLL+NNLO
      prediction of \HQT{} (green) for the Higgs transverse momentum.
      In \HQT{} we keep the resummation scale $Q_{\rm res}$ always
      fixed to $m_H/2$ and vary $\mu_R$ and $\mu_F$.  On the left
      (right), the NNLOPS (\HQT{}) uncertainty band is shown. In the
      lower panel, the ratio to the NNLOPS (\HQT{}) central
      prediction is displayed.}
    \label{fig:pth}
\end{figure}

Fig.~\ref{fig:pth} shows the Higgs transverse momentum $p_T^H$. We
compare our simulation with
\HQT{}~\cite{Bozzi:2005wk,deFlorian:2011xf}, whose central value is
obtained with $Q_{\rm res}=m_H/2$ and $\mu_R=\mu_F=m_H/2$. The \HQT{}
result corresponds to a NNLL prediction for $p_T^H$, matched to the
fully inclusive cross section at NNLO. Here we notice that the two
results are almost completely contained within each other's
uncertainty band in the region of low-to-moderate transverse
momenta. The central values at small momenta also exhibit a very good
agreement, supporting our choice for $\beta$. The difference in the
large-$p_T$ tail is not a reason of concern, and it is expected since
the two predictions use different scales at large $p_T$, as explained
in ref.~\cite{Hamilton:2013fea}.

%% \begin{figure}[!htb]
%%   \begin{center}
%%     \includegraphics[scale=0.6]{plots/NNLOPS/fig_5/fig5L.pdf}~
%%     \includegraphics[scale=0.6]{plots/NNLOPS/fig_5/fig5R.pdf}
%%     \caption{Comparison of the NNLOPS (red) with the NNLL+NNLO
%%       prediction of \HQT{} (green) for the Higgs transverse momentum.
%%       In \HQT{} we keep the resummation scale $Q_{\rm res}$ always
%%       fixed to $m_H/2$ and vary $\mu_R$ and $\mu_F$.  On the left
%%       (right), the NNLOPS (\HQT{}) uncertainty band is shown. In the
%%       lower panel, the ratio to the NNLOPS (\HQT{}) central
%%       prediction is displayed.}
%%     \label{fig:pth}
%%   \end{center}
%% \end{figure}

Finally, we also mention that a comparison among NNLOPS and NNLL+NNLO
predictions from \JETVHETO{}~\cite{Banfi:2012jm} was successfully
carried out for the jet veto efficiency, defined as the cross section
for producing the Higgs boson and no jets with transverse momentum
greater than a given value ($p_{\scriptscriptstyle \mathrm{T,veto}}$),
divided by the respective total inclusive cross section. The central
predictions of the two programs are never out of agreement by more
than 5-6\%, and the two sets of predictions lie within each other's
error bands essentially everywhere over all values of
$p_{\scriptscriptstyle \mathrm{T,veto}}$, as shown in
ref.~\cite{Hamilton:2013fea}.

\section*{Acknowledgments}
NNLOPS results presented here have been obtained in
ref.~\cite{Hamilton:2013fea}, in collaboration with K.~Hamilton,
P.~Nason and G.~Zanderighi. The original proposal of reaching NNLOPS
accuracy from \MiNLO{}-merged NLOPS simulations was outlined in
ref.~\cite{Hamilton:2012rf}, which was co-authored by C.~Oleari. The
author acknowledges the organising committee for the financial help
received to cover the living expenses.

\end{document}